\documentclass[12pt]{iopart}

\usepackage{graphicx}
 \usepackage{comment}
\usepackage{color}
\usepackage{hyperref}
\usepackage{soul}

\usepackage{iopams}

\expandafter\let\csname equation*\endcsname\relax

\expandafter\let\csname endequation*\endcsname\relax

\usepackage{amsmath}

\hypersetup{colorlinks=true, linkcolor=blue, citecolor=blue, urlcolor=blue}

\begin{document}

\title[Collective excitations and supersolid behavior...]{Collective excitations and supersolid behavior  \\of bosonic
  atoms inside two crossed optical cavities}

\author{J. Lang$^1$, F. Piazza$^2$, W. Zwerger$^1$}
\address{$^1$Physik Department, Technische Universit\"at M\"unchen, 85747 Garching, Germany\\
$^2$Max-Planck Institute for the Physics of Complex Systems, N\"othnitzer Stra{\ss}e 38, 01187 Dresden, Germany }
\ead{j.lang@tum.de}

\begin{abstract}
We discuss the nature of symmetry breaking and the associated
collective excitations for a system of bosons coupled to the electromagnetic 
field of two optical cavities. For the specific configuration realized in a recent experiment at 
ETH~\cite{leonard2016supersolid,leonard_supersolid_goldstone}, we show
that, in absence of direct intercavity scattering and for parameters chosen such
that the atoms couple symmetrically to both cavities, the system possesses an
approximate $U(1)$ symmetry which holds asymptotically for
vanishing cavity field intensity. It corresponds to the
invariance with respect to redistributing the total intensity
$I=I_1+I_2$ between the two cavities. The spontaneous
breaking of this symmetry gives rise to a broken continuous translation-invariance for the
atoms, creating a supersolid-like order in the presence of a Bose-Einstein condensate. In particular, we show 
that atom-mediated scattering between the two cavities, which favors the state with equal light intensities $I_1=I_2$ 
and reduces the symmetry to $\mathbf{Z}_2\otimes \mathbf{Z}_2$, gives rise to a finite value $\sim \sqrt{I}$ of the 
effective Goldstone mass. For strong atom driving, this low energy mode is clearly separated from 
an effective Higgs excitation associated with changes of the total intensity $I$.  
In addition, we compute the spectral distribution of the cavity light field and show that 
both the Higgs and Goldstone mode acquire a finite lifetime due to Landau damping at non-zero temperature. 
\end{abstract}

\maketitle

\section{Introduction}
{\color{red}}

The notion of a supersolid, i.e. a solid which is able to sustain dissipationless mass currents 
typical for superfluids, is clearly highly counterintuitive~\cite{boninsegni2012}. It requires that
the particles in the solid can effectively move freely through quantum mechanical delocalization~\cite{yang62}.
A conceptually simple example, suggested by Andreev and Lifshitz~\cite{andreev1969} and by Chester~\cite{chester1970},
is a quantum crystal with a finite density of defects even at zero temperature. With Bose statistics, the resulting
dilute gas of defects is expected to undergo BEC at low temperatures, giving rise to a finite superfluid density 
and thus e.g. to a reduced moment of rotational inertia~\cite{leggett1970}. As shown by Prokof'ev and 
Svistunov~\cite{prokofev2005}, this scenario for supersolidity is in fact the generic one because superfluid 
states are necessarily gapless with respect to adding and removing particles. A commensurate supersolid with 
an integer number of particles in the unit cell, in turn, requires a fine tuned value of the density. Formally, 
such a state appears in the superfluid regime of the Bose-Hubbard model~\cite{fisher1989} at 
fixed integer densities. While such a commensurate superfluid can in principle be realized with ultracold atoms 
in an optical lattice~\cite{greiner2002},  it is important to stress that - similar to the case of superfluid Helium on the surface of a 
regular crystal - this is not a supersolid in the usual sense because translation invariance is broken 
{\it externally} and not spontaneously as a result of the interactions between the particles. This differs 
from several proposed realizations with ultracold atoms, where dipolar~\cite{giovannazzi_dipolar_2002} or Rydberg~\cite{henkel_rydberg_2010} 
interactions as well as collective light scattering~\cite{ostermann_CARL_2016} or
spin-orbit coupling~\cite{martone_SOC_2013} give rise to crystalline order. In particular, in the latter context, the recent 
observation of a stripe phase at MIT~\cite{MIT_stripeSS_2017} provides a simple example of supersolid-like order. \\ 

A quite different system in which long range positional order may coexist with superfluid behavior 
has been realized in recent years by studying ultracold atoms in a high finesse cavity. In the 
presence of a transverse laser field there is an induced interaction between the atoms 
which is mediated by the cavity photons~\cite{cavity_rmp}. The interaction is long-ranged
and, beyond a critical strength $\lambda_c$ of the drive, the atoms spontaneously arrange in a periodic lattice,
allowing to scatter the light from the transverse field coherently into the cavity~\cite{baumann2010,black2003}.
This is an example of the classic Dicke-Hepp-Lieb transition to a superradiant state~\cite{dicke1954,hepp1973,wang1973,nagy_2010}
and it results in a two-fold degeneracy of the periodic arrangement of the atoms. More precisely, 
the $\mathbf{Z}_2$ symmetry which is broken at the Dicke-Hepp-Lieb transition is associated 
with the relative sign of the two degenerate wave-vectors $\mathbf{q}=\pm\mathbf{k}_0$ which appear with 
equal weight in the standing periodic density wave described by a non-vanishing 
expectation value $\langle\hat{\rho}_{\mathbf{k}_0}\rangle\ne 0$ of the density operator $\hat{\rho}_{\mathbf{q}}$ 
($\mathbf{k}_0$ is the externally fixed cavity wave vector). From the point of view of 
off-diagonal long range order, which characterizes BEC in its most general form~\cite{penrose1956}, 
the phase beyond $\lambda_c$ is one in which extensive eigenvalues of the one particle density matrix appear not only at 
$\mathbf{q}=0$ but also at arbitrary multiples of $\mathbf{q}=\pm\mathbf{k}_0$, forming a fragmented condensate
with a self-generated optical lattice~\cite{piazza_bose,lode_DHL_2017}.
The system therefore possesses simultaneously both diagonal and off-diagonal long range order. 
Despite the fact that periodic order is now generated through light-field mediated interactions between 
the atoms, it is not a supersolid in the standard sense because it does not sustain dissipationless particle 
currents e.g. of the $\mathbf{q}=0$ part of the condensate with respect to the fixed periodic density wave
pattern associated with the $\pm\mathbf{k}_0$ components
\footnote{ Within a hydrodynamic description, such dissipationless currents would be associated 
with a fourth sound-like mode with linear dispersion, see~\cite{liu1978}}.    
Moreover, due to the long range nature of the 
interaction the system is effectively zero-dimensional and there are no proper Goldstone 
modes usually associated with the breaking of a continuous translation symmetry, which are the
acoustic phonons near reciprocal lattice vectors $\mathbf{q}\simeq \mathbf{G}$~\cite{wagner1966}.\\

Recently, a major step towards the realization of supersolid behavior with dissipationless particle transport 
has been taken by L\'eonard and coworkers at ETH in a setup involving ultracold atoms 
in \emph{two crossed} cavities~\cite{leonard2016supersolid}. In this setup,
a cloud of Bose-condensed atoms is enclosed in a configuration involving two optical cavities which
are at a $60^\circ$ angle with respect to each other. Tuning the parameters such that the atoms
couple symmetrically to both cavities, this system allows to realize light-induced crystallization of
the atoms which involves an arbitrary superposition of both cavity wave vectors. Within a simple 
two-mode description, the two discrete symmetries $\mathbf{Z}_2$ of the individual cavities can 
thus be combined to a continuous $U(1)$ symmetry, allowing to observe a continuous shift of 
the crystallization pattern~\cite{leonard2016supersolid}. 
Our aim in the present work is to 
analyze a fully microscopic model for this setup in order to study the detailed 
structure of the broken symmetries and the resulting spectrum of collective 
excitations. In particular, we will derive the associated effective dynamic Ginzburg-Landau functional 
for the light field in the cavity and discuss the limits in which the system indeed exhibits
the breaking of a continuous translation symmetry. Beyond a detailed discussion of 
symmetry breaking and the subtle issue of supersolidity in this context, our results also 
provide a quantitative understanding of the recent measurements of the effective 
Goldstone and Higgs mode frequencies~\cite{leonard_supersolid_goldstone}.

\section{Model and Symmetries}
\label{sec:symmetries}

\begin{figure}[htp]
\centering
\includegraphics [width=0.7\textwidth]{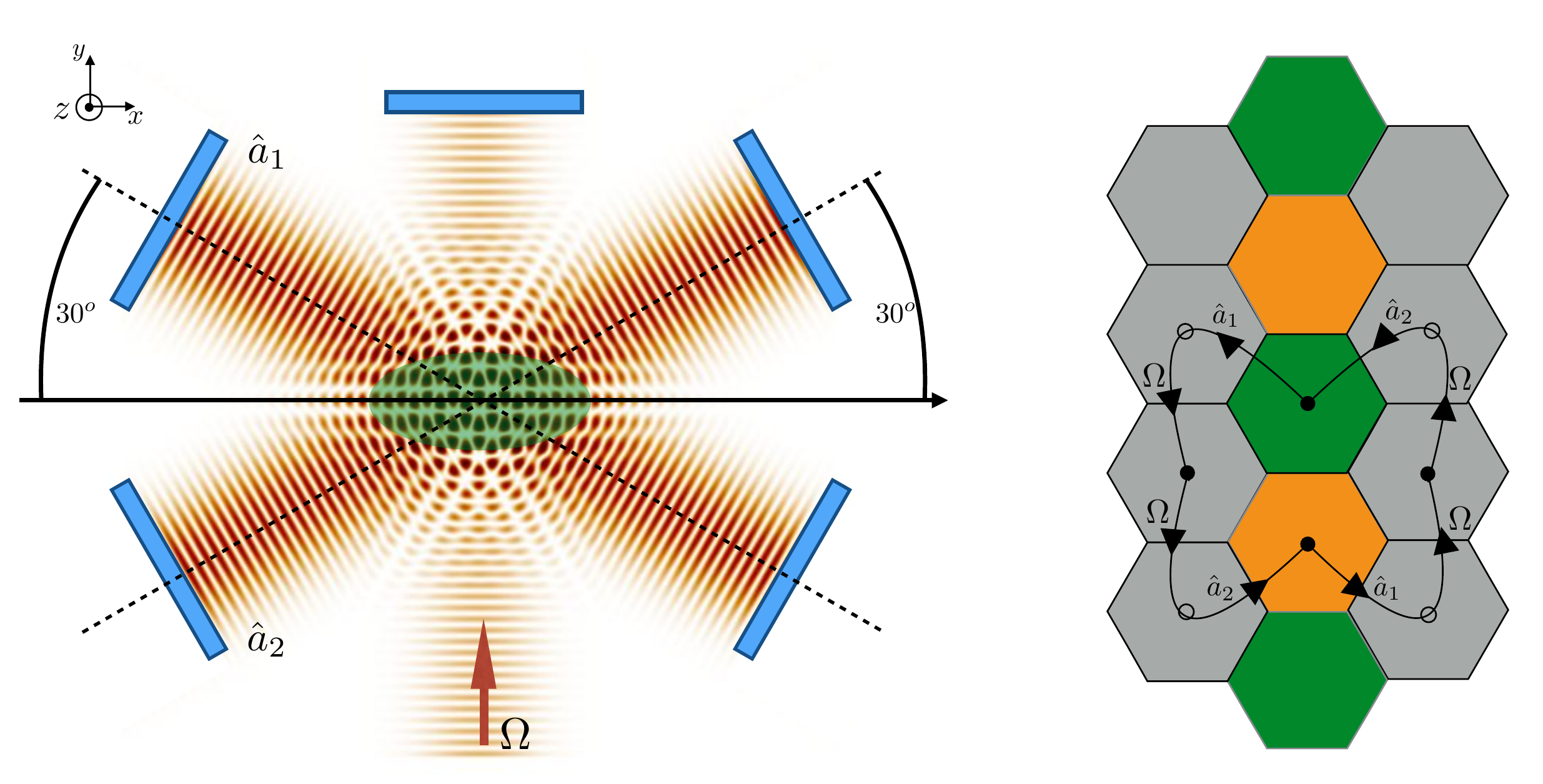}
\caption{Left: sketch of the setup considered as implemented in the
  experiments of \cite{leonard2016supersolid}. Right: momentum space
  in the repeated-zone scheme, where each hexagon indicates the
  Brillouin zone for a given band. Green zones are occupied even
  without cavity fields and in particular their center
  ($\Gamma$-point) is the only state occupied at $T=0$. Gray zones are
occupied via cavity-photon scattering and correspond to the truncation
used in the $U(1)$-symmetric Hamiltonian (\ref{eq:Htrunc}). The closed
curve indicates a scattering path involving two photons from each
cavity. This scattering process, for which we need to include the
orange zones in our Hilbert-space truncation, explicitly breaks the $U(1)$ symmetry
of the full Hamiltonian (\ref{eq:ham}).}
\label{fig:setup}
\end{figure}

We consider the setup used for the recent experiments at ETH
\cite{leonard2016supersolid,leonard_supersolid_goldstone}. It
consists of a three dimensional cloud of bosonic atoms trapped at the intersection of the
TEM$_{00}$-modes of two optical cavities. All photons couple the atomic ground state to the same excited state. 
Generalizing the formalism developed in a previous paper~\cite{piazza_bose}, adiabatic elimination of the excited state (which is well justified for the experimental setup) leads to an effective Hamiltonian. In the frame rotating with the driving laser $\omega_L$ it takes the following form (note that we use units in which $\hbar=1$ throughout the paper):  
\begin{align}
\label{eq:ham}
\hat{H}=-\sum_{i=1,2}\Delta_i\hat{a}_i^\dag \hat{a}_i+\int d\mathbf{r}\; \hat{\psi}^\dag(\mathbf{r})\left(-\frac{\nabla^2}{2m}+\hat{V}(\mathbf{r})\right)\hat{\psi}(\mathbf{r})\; .
\end{align}
Here, $\hat{a}_i$ is the bosonic annihilation operator of a photon in the
cavity $i$, $\hat{\psi}(r)$ is the bosonic annihilation operator for an
atom at position $\mathbf{r}$ with mass $m$ and $\Delta_i=\omega_L-\omega_i<0$ is the detuning of the laser from the cavity resonance.
Assuming equal dipole couplings $g_1=g_2=g$ in both cavities, 
the associated single-particle potential---which still depends on the quantum state of the cavity field---is given by 
\begin{align}
\label{eq:pot_2cav}
\hat{V}(\mathbf{r})=V_L(y)+\sum_{i=1,2} V_i(\mathbf{r})\left(\hat{a}_i+\hat{a}^\dag_i\right)+\sum_{i,j=1,2}V_{i,j}(\mathbf{r})\hat{a}^\dag_i\hat{a}_j\;,
\end{align}
where $V_L$ accounts for the pump potential while $V_{1}$ and $V_{2}$ are 
the potentials resulting from the interference between one cavity and the pump. They are given by  
\begin{align}
V_L&=\frac{\Omega^2}{\Delta_A}\cos^2(k_0y+\frac{\pi}{2})\notag\\
V_1&=g\frac{\Omega}{\Delta_A}\cos(k_0y+\frac{\pi}{2})
  \cos(\frac{\sqrt{3}}{2}k_0x+\frac{1}{2}k_0y)\notag\\
V_2&=g\frac{\Omega}{\Delta_A}\cos(k_0y+\frac{\pi}{2})
  \cos(\frac{\sqrt{3}}{2}k_0x-\frac{1}{2}k_0y)\;,
\end{align}
where $\Omega$ is the Rabi frequency of the driving laser while 
$\Delta_A=\omega_L-\omega_A<0$ is the detuning of the atomic resonance from the 
driving laser at frequency $\omega_L$. Note that in the experiment under consideration \cite{leonard2016supersolid,leonard_supersolid_goldstone} the laser drive is far detuned ($\Omega/|\Delta_A|\approx 3*10^{-4}$), which justifies the adiabatic elimination underlying the effective Hamiltonian \eqref{eq:ham}. 
The last term in Eq.~(\ref{eq:pot_2cav}) describes the effects of direct inter- and 
intracavity scattering. The associated effective potential $V_{i,j}(\mathbf{r})=\eta_i(\mathbf{r})\eta_j(\mathbf{r})$
is determined by the two mode functions
\begin{align}
\eta_1(\mathbf{r})=\frac{g}{\sqrt{|\Delta_A|}}\cos(\frac{\sqrt{3}}{2}k_0x+\frac{1}{2}k_0y)\;\;\; {\rm and} \;\;\; 
\eta_2(\mathbf{r})=\frac{g}{\sqrt{|\Delta_A|}}\cos(\frac{\sqrt{3}}{2}k_0x-\frac{1}{2}k_0y)\;,
\end{align}
which correspond to the configuration shown in Fig.~\ref{fig:setup}, where
the cavity axes form a $30^\circ$ angle with the $x$ axis and lie in
the $x-y$ plane i.e. $\mathbf{k}_{1,2}=k_0(\mathbf{n}_x
\cos(30^\circ)+\mathbf{n}_y \sin(30^\circ))$. The pump axis is
along the $y$ direction i.e. $\mathbf{k}_p=k_0 \mathbf{n}_y $ and the standing wave
obtained by retroreflection has a phase-shift $\pi/2$. Due to the small detuning $|\Delta_i|\ll\omega_L$ the pump and cavity modes can be taken to have the same wavelength $\lambda_0=2\pi/k_0$.
In the following discussion we will include the leading contribution 
of the direct intracavity processes $\propto \eta_i(\mathbf{r})\eta_i(\mathbf{r})$ as a dispersive shift to the cavity detuning
\begin{align}
\delta_{c_i}=-\Delta_i+\frac{N g^2}{2 \Delta_A}>0\;.
\end{align}
We will however neglect contributions quadratic in $V_{i,j}(\mathbf{r})$, which are of order $g^4$. This is valid in the experimentally realized regime
where fourth order processes due to $V_1$ and $V_2$ are more important than second order effects in $V_{i,j}$. Since intermediate states in this perturbation series carry energies $\sim E_R$ (in the superradiant phase creation of cavity photons costs very little energy), this reduces to the condition $\Omega\gg \sqrt{|\Delta_A E_R|}$ with recoil energy $E_R=k_0^2/2m$, which is well satisfied in the experiment (see also below). As the critical Rabi amplitude $\Omega_c \propto |\Delta_A|$ is decreased, close to the onset of superradiance, direct intercavity scattering will eventually be the dominant effect. The role of these processes, together with different choices
of the retroreflection phase-shift, have been theoretically
investigated in \cite{safei_2015,vestigial_order}.

As discussed in \cite{leonard2016supersolid}, assuming
small (in a sense that will become clear later) light
field intensities so that multiple scattering is suppressed, we can
restrict the Hilbert space to the lowest nine momentum states
$|k_x,k_y\rangle=|0,0\rangle,|\pm\mathbf{k}_i\pm\mathbf{k}_p\rangle$
(see also Fig.~\ref{fig:setup})
and truncate the Hamiltonian (\ref{eq:ham}) as follows
\cite{leonard2016supersolid}:
\begin{align}
\label{eq:Htrunc}
\hat{H}_{\rm trunc}=\sum_{i=1,2}\left[\delta_{c_i}\hat{a}_i^\dag \hat{a}_i+E_+\hat{c}_{i+}^\dag \hat{c}_{i+}+E_-\hat{c}_{i-}^\dag \hat{c}_{i-}-\frac{g\Omega}{2\sqrt{2}\Delta_A}(\hat{a}_i^\dag+\hat{a}_i)(\hat{c}_{i+}^\dag\hat{c}_0+\hat{c}_{i-}^\dag\hat{c}_0+\text{h.c.})\right]\;,
\end{align}
where $\hat{c}_{i\pm}^\dag$ excites an atom into a standing wave of
wave vector $\mathbf{k}_i\pm\mathbf{k}_p$ with energy
$E_{\pm}=(2\pm1)E_R$ and $\hat{c}_0$ removes an atom at $\mathbf{k}=0$. Here
$g\Omega/|\Delta_A|$ is the effective cavity pump strength.
The truncated Hamiltonian (\ref{eq:Htrunc}) has in general a
$\mathbf{Z}_2\otimes \mathbf{Z}_2$ symmetry corresponding to the following transformation
\begin{align}
\label{eq:Z2trafo_trunc}
(\hat{a}_1,\hat{c}_{1\pm})&\to-(\hat{a}_1,\hat{c}_{1\pm})\\
(\hat{a}_2,\hat{c}_{2\pm})&\to-(\hat{a}_2,\hat{c}_{2\pm})\;.
\end{align}
The spontaneous breaking of either one of these discrete symmetries corresponds to a
superradiant phase transition characterized by the order parameter
$\langle\hat{a}_i^\dag+\hat{a}_i\rangle$ or equivalently
$\langle \hat{c}_{i+}^\dag\hat{c}_0+\hat{c}_{i-}^\dag\hat{c}_0+\text{h.c.}\rangle$. In the full
model in real space given by Eq.~(\ref{eq:ham}), the above
$\mathbf{Z}_2\otimes \mathbf{Z}_2$ symmetry corresponds to the transformations
\begin{align}
\label{eq:Z2trafo}
(\hat{a}_1,\mathbf{r})&\to(-\hat{a}_1, \mathbf{r}+\pi\frac{\mathbf{k}_1}{|\mathbf{k}_1|^2})\\
(\hat{a}_2,\mathbf{r})&\to(-\hat{a}_2, \mathbf{r}+\pi\frac{\mathbf{k}_2}{|\mathbf{k}_2|^2})\;,
\end{align}
which involve a discrete spatial translation along a cavity axis. In
this sense, the superradiant transition corresponds to a self-ordering
of the atoms into a spatial pattern which scatters constructively into
the cavity \cite{cavity_rmp}. 

As pointed out in \cite{leonard2016supersolid}, for a symmetric choice
of cavity detunings $\Delta_1=\Delta_2$ there is an accidental
$U(1)$ symmetry in the truncated Hamiltonian (\ref{eq:Htrunc}):
\begin{align}
\label{eq:U1trafo_trunc}
&\big(\hat{a}_1,\hat{c}_{1\pm},\hat{a}_2,\hat{c}_{2\pm}\big)\notag\\ &\to\big(\hat{a}_1\cos{\theta}-\hat{a}_2\sin{\theta},\hat{c}_{1\pm}\cos{\theta}-\hat{c}_{2\pm}\sin{\theta},\hat{a}_1\sin{\theta}+\hat{a}_2\cos{\theta},\hat{c}_{1\pm}\sin{\theta}+\hat{c}_{2\pm}\cos{\theta}\big)\;.
\end{align}
The signatures of the spontaneous breaking of this continuous symmetry, which corresponds to a fixed value of the relative phase $\theta$ of
the two coherent cavity fields, varying randomly between different realizations, have been experimentally investigated in
\cite{leonard2016supersolid,leonard_supersolid_goldstone}. In particular, it has been shown that the symmetry broken state possesses a collective excitation with a frequency below the experimental resolution of $100$kHz. Correspondingly the cavity fields have been observed at randomly distributed relative amplitudes with a fixed overall output intensity. Both these signatures have been interpreted as the Goldstone mode of the broken $U(1)$ symmetry.

As discussed in~\cite{leonard2016supersolid}, the transformations (\ref{eq:U1trafo_trunc}) can be translated
 into an invariance of the potential (\ref{eq:pot_2cav}) in real space. Indeed,  
by examining the potential \eqref{eq:pot_2cav} we see that by restricting
to the subspace $X$ defined by $k_0y+\pi/2=n\pi,n\in\mathbb{Z}$, the
potential is invariant under:
\begin{align}
\label{eq:U1trafo}
&\big (\hat{a}_1,\hat{a}_2,x,y=\pi(n-1/2)/k_0\big)\notag\\
&\to\big(\hat{a}_1\cos{\theta}-\hat{a}_2\sin{\theta},\hat{a}_1\sin{\theta}+\hat{a}_2\cos{\theta}, x\pm2\theta/(\sqrt{3}k_0),y=\pi(n-1/2)/k_0\big)\;,
\end{align}
where the $-(+)$ sign applies for even (odd) values of $n$.
The continuous symmetry of the Hamiltonian under rotations of the cavity field by an angle $\theta$ and a simultaneous shift of the atoms along the $x$-direction by $\pm2\theta/\left(\sqrt{3}k_0\right)$ thus leads to supersolid-like behavior with no restoring force for translations of the atoms along the $x$-direction.

Now, the fact that the $U(1)$ symmetry
(\ref{eq:U1trafo}) in the full model (\ref{eq:ham}) is restricted
to the subspace $X$ of discretely spaced values of the $y$ coordinate
implies that this symmetry holds only approximately.
The fundamental reason is that the potential (\ref{eq:pot_2cav}) has no
minimum on the $U(1)$-symmetric lines $y=\pi(n-1/2)/k_0$, but rather
at a position which is shifted by an amount inversely proportional to
the amplitude of the state independent ac-Stark shift $V_L$ in the effective potential. This shift appears due to interference between the two
cavity fields and is therefore present for any finite number of photons in
\emph{both} cavities. The corresponding lowest-order scattering processes are
depicted in Fig.~\ref{fig:setup} as a closed path involving two
photons for each cavity, which for equal intensities $I_1=I_2=I$ implies that the
lowest order of the explicit breaking of the $U(1)$ symmetry is proportional to $I^2$. It is important to note that the description of these scattering
processes requires the inclusion of momentum states that are absent in
the truncation used to obtain the Hamiltonian (\ref{eq:Htrunc}) (see Fig.~\ref{fig:setup}).
In the following we will discuss the consequences of the explicit $U(1)$
symmetry breaking for the supersolid-like features, which will turn
out to be still approximately present in the limit of intense laser
driving $\Omega$.

To understand the physics beyond the deviations from a perfect $U(1)$-symmetry, it is convenient to use 
a simple effective Hamiltonian obtained by adiabatically eliminating the photons from \eqref{eq:ham}.
Assuming deep lattices such that we can neglect the kinetic
term as well as all terms beyond $\mathcal{O}(g^2)$ from the contribution $V_{i,j}(\mathbf{r})$, the resulting effective Hamiltonian 
\begin{align}
\hat{H}=\int d\mathbf{r} \;\hat{\psi}^\dag(\mathbf{r})\hat{\psi}(\mathbf{r})\left(V_L(\mathbf{r})-\sum_i \frac{V_i(\mathbf{r})}{\delta_{c_i}}\int d\mathbf{r'}\; V_i(\mathbf{r'})\hat{\psi}^\dag(\mathbf{r'})\hat{\psi}(\mathbf{r'})\right)
\end{align}
for the atoms alone contains an instantaneous, cavity field induced, attractive density-density interaction of 
the form $-\sum_i V_i(\mathbf{r}) V_i(\mathbf{r'})/\delta_{c_i}$ which does not decay as a function of the separation $|\mathbf{r}-\mathbf{r'}|$. 
Since we neglect direct intercavity scattering, there are no interactions of higher order in the density.
In the case that only a single cavity is superradiant the ground state is given by a density distributed solely within the high symmetry subspace $X$
\begin{align}
\label{eq:densGS1}
\rho(\mathbf{r})=\rho_0\frac{2\pi^2}{\sqrt{3}}\delta\left(k_0 y-\frac{\pi}{2}+n\pi\right)\delta\left(k_0 x-\frac{\mp\pi+4\pi m}{2\sqrt{3}}\right),~n,m\in\mathbb{Z}
\end{align}
with $\rho_0=N/V$. Here, the minus sign implies superradiance in cavity 1 ($\alpha_1=\langle \hat{a}_1\rangle\neq0$) while the plus sign corresponds to a finite expectation value of $\alpha_2=\langle \hat{a}_2\rangle$. The energy density of both states is given by $\epsilon=\frac{\Omega^2}{\Delta_A}(1+c)$, with 
$c=N g^2/\delta_{c_i}|\Delta_A|$ a dimensionless positive constant  which is much less than one for typical experimental parameters. 
For two identical cavities $\delta_{c_i}=\delta_c$ and therefore $\alpha=\alpha_i$, this is, however, not in the ground state manifold which instead contains for example the density profile
\begin{align}
\label{eq:densGS2}
\rho(\mathbf{r})=\rho_0\frac{2\pi^2}{\sqrt{3}}\delta(k_0 x -\frac{2 n \pi}{\sqrt{3}})\sum_{\sigma=\pm}\delta\left(k_0 y +2\sigma \arcsin\left(\sqrt{\frac{1+4 c-d(c)}{6 c}}\right)+2\pi m\right),
\end{align}
with $n,m \in \mathbb{Z}$ and $d(c)=\sqrt{1+2 c+4c^2}$.
This density distribution slightly frustrates the potential $V_L$ induced by the ac-Stark shift of the atoms
and shifts the densities away from the $X$ subspace. It
therefore slightly breaks $U(1)$ invariance in the atomic density and
locks the relative cavity phases. The small energy difference between
state (\ref{eq:densGS2}) and state (\ref{eq:densGS1}) is  given by
\begin{align}
\label{eq:energyshift}
&\Delta \epsilon=-\frac{\Omega^2}{\Delta_A}\bigg[1+c
+\frac{1}{27c^2}\left(1+4 c-d(c)\right)\left(1-2 c -d(c)\right)
\left(2+2c+d(c)\right)\bigg]\approx\frac{\Omega^2 c^2}{4 \Delta_A}<0\;.
\end{align}
As will be discussed below, this energy determines the scale of the effective Goldstone mass. 
Resubstituting either one of these density profiles into the cavity equations of motion we obtain
$|\alpha|=\sqrt{I}\approx N\Omega g /|\Delta_A|\delta_c$ to leading order in $c$,
which shows that in the deep lattice limit the critical coupling
strength $\lambda_c$ vanishes.  Since kinetic energy contributions have been neglected, 
the Goldstone mass has an upper bound
\begin{align}
\label{eq:mG}
m_G =\frac{\sqrt{-\Delta \epsilon}}{\alpha}\lesssim \frac{\delta_c\sqrt{-\Delta_A I}}{\Omega N}=\frac{g}{\sqrt{-\Delta_A}}\;.
\end{align}
Physically the Goldstone mass $m_G$ arising from the finite energy scale 
$\Delta \epsilon$ associated with the breaking of the $U(1)$ symmetry
describes the azimuthal curvature of the Ginzburg-Landau
potential, which will be discussed in more detail in section~\ref{sec:GLpotential}.
As expected according to the argument based on the scattering processes
illustrated in Fig.~\ref{fig:setup}, $\Delta\epsilon\propto |\alpha^4|=I^2$.
The explicit symmetry breaking $\Delta\epsilon$ due to the latter
scattering processes has actually the same scaling with intensity as
the one which would result from direct intercavity scattering (not
involving the pump $\Omega)$, which we neglected in our model
(\ref{eq:ham}). As mentioned before, this is justified in the
experimentally realized limit $\Omega/\sqrt{|\Delta_A E_R|}\gg 1$, where direct
intercavity scattering is suppressed with respect to the processes
shown in Fig.~\ref{fig:setup}. In particular, the fact that we can
neglect all contributions from the last term in \eqref{eq:pot_2cav}
beyond the simple dispersive shift has no influence on the $U(1)$
invariance in subspace $X$. 
Even including all contributions from the last term in \eqref{eq:pot_2cav}, the explicit breaking of the
$U(1)$ symmetry is still caused by the fact that the global potential minimum
for the atoms lies outside the subspace $X$.

\section{Effective action and phase diagram}
\label{sec:formalism}

In order to compute the phase diagram and the experimentally accessible
spectrum of the cavity light field, we extend the effective equilibrium path-integral approach
developed in \cite{piazza_bose} for a single-cavity configuration. We
derive an effective action for the cavity degrees of freedom by
exactly integrating out the atoms. The action
splits into a mean-field (MF) plus a fluctuation (FL) part. The latter will be discussed in
detail in section \ref{sec:low-energy-fluctuations} below.  As shown in \cite{piazza_bose}, this
action becomes exact in the thermodynamic limit due to the infinite-range
interactions. By separating the coherent part of cavity fields
$\alpha_i=\langle\hat{a}_i\rangle$ as well as the atom field
$\phi(\mathbf{r})=\langle\hat{\psi}(\mathbf{r})\rangle$ (which corresponds to the condensate fraction) we obtain
the effective action 
\begin{align}\label{eq:S_eff}
S_{\rm  eff}[a_{1,2}^*,a_{1,2}]=S_{\rm  eff}^{(MF)}+S_{\rm  eff}^{(FL)}[a_{1,2}^*,a_{1,2}]\;.
\end{align}
The leading MF action reads
\begin{align}
\label{eq:S_MF}
S_{\rm
  eff}^{(MF)}=-\sum_{i=1,2}\Delta_i|\alpha_i|^2+\mathrm{Tr}\ln\left[G^{-1}\right]+\int
d\mathbf{r}\;\phi^*(\mathbf{r})\left[-\frac{\nabla^2}{2m}+V_{\rm sp}(\mathbf{r})-\mu \right]\phi(\mathbf{r})\, .
\end{align}
It involves an effective c-number single-particle potential $V_{\rm sp}(\mathbf{r})=\hat{V}(\mathbf{r})\bigg|_{\hat{a}_i\to\alpha_i}$
felt by the atoms in which the light field operators are replaced by their coherent state expectation values.  
The atom propagator defined by
\begin{align}
G_{n,n'}^{-1}(\mathbf{r,r'})=\left[-i\omega_n-\frac{\nabla^2}{2m}+V_{\rm
  sp}(\mathbf{r})-\mu\right]\delta_{n,n'}\delta(\mathbf{r-r'})\;,
\end{align}
where the integers $n,n'$ label the Matsubara space with frequencies $\omega_n=2\pi n
k_BT$, $k_B$ is the Boltzmann constant and $T$ the temperature
of the system. In Eq.~\eqref{eq:S_MF} the trace  $\mathrm{Tr}=\int
d\mathbf{r}\sum_n$ is taken over coordinate and Matsubara space and $\mu$ is
the atomic chemical potential. 

The saddle-point associated with the mean-field action defines a
closed system of equations:
\begin{align}
\label{eq:mf}
-\Delta_{i}\alpha_{i}+\int
d\mathbf{r}\;\frac{\partial V_{\rm
    sp}(\mathbf{r})}{\partial\alpha_{i}^*}\left[|\phi(\mathbf{r})|^2+\rho(\mathbf{r})\right]&=0\;,\;\;i=1,2\notag\\
\left(-\frac{\nabla^2}{2m}+V_{\rm
  sp}(\mathbf{r})-\mu \right)\phi(\mathbf{r})&=0\notag\\
\int d\mathbf{r}\;\left(|\phi(\mathbf{r})|^2+\rho(\mathbf{r})\right)&=N\;,
\end{align}
with the condensed $\rho_0(\mathbf{r})=|\phi(\mathbf{r})|^2$ and non-condensed
$\rho(\mathbf{r})=\langle \hat{\psi}^\dag(\mathbf{r}) \hat{\psi}(\mathbf{r})\rangle-\rho_0(\mathbf{r})=\sum_\ell n_{\rm
  b}(\epsilon_\ell-\mu)|v_\ell(\mathbf{r})|^2$ atom density. Here
$v_\ell(\mathbf{r})$ are the eigenvectors of the single-atom
Hamiltonian with potential $V_{\rm sp}(\mathbf{r})$ with eigenvalue
$\epsilon_{\ell}$, and $n_b(x)=(\exp(x/k_BT)-1)^{-1}$ is the Bose-Einstein distribution.
The second equation in \eqref{eq:mf} is the
Gross-Pitaevskii equation for the condensate wavefunction while the
third equation fixes the chemical potential.
We stress that both $\rho(\mathbf{r})$ and $V_{\rm
  sp}(\mathbf{r})$ depend on the cavity coherent parts $\alpha_{1,2}$.\\
It is convenient to introduce dimensionless quantities, which we define by 
\begin{align}
\delta_c=\min_{i}\left(\delta_{c_i}\right),~n=\frac{\rho_0}{\left(m \delta_c\right)^{3/2}},~U_p=\frac{\Omega^2}{|\Delta_A| \delta_c},~\lambda=\frac{\Omega g \sqrt{N}}{|\Delta_A|\delta_c},~\epsilon_R=\frac{E_R}{\delta_c}\;.
\end{align}
Moreover, we measure temperatures in units of the critical temperature of an ideal Bose gas 
$T_c^\text{ideal}=2\pi\delta_c\left(n/\zeta(3/2)\right)^{2/3}$ at the given average density $n$ 
and with $\zeta$ the Riemann zeta function. We furthermore rescale $\alpha \rightarrow \alpha/\sqrt{N}$ in the remainder of this paper and in all figures.

\begin{figure}[t]
\centering
\includegraphics [width=0.55\textwidth]{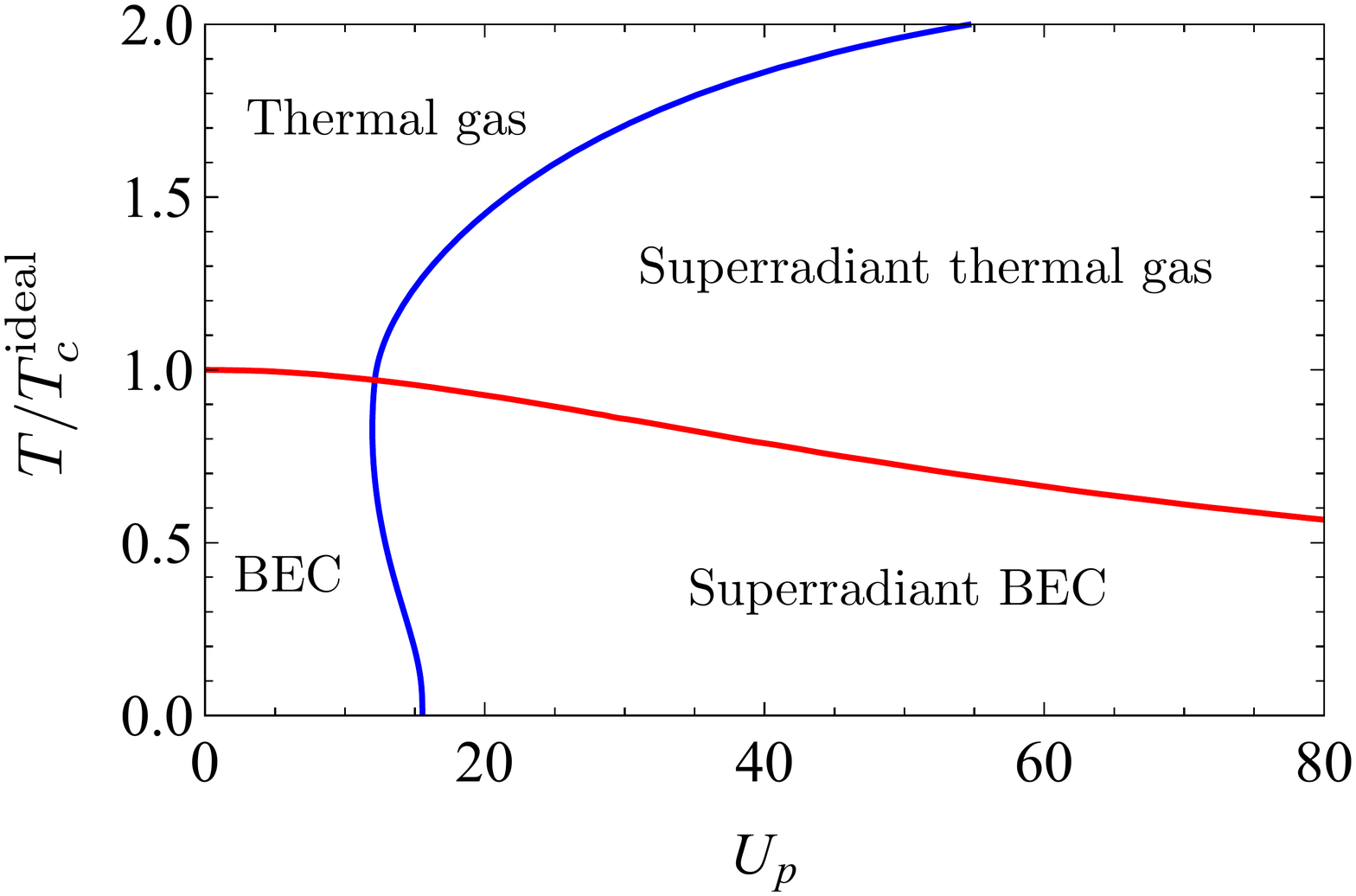}
\caption{Phase diagram in the $U_p-T$ plane for parameters $\Delta_2=\Delta_1$, $\epsilon_\text{R}=8$, $n=1$, $\lambda=2.8$. To the right of the blue line the system is in a superradiant state with equal intensity in both cavities, below the red line a finite fraction of the atoms is condensed.}
\label{fig:etaT_PD}
\end{figure}

As discussed in the previous section, the Hamiltonian \eqref{eq:ham}
possesses the $\mathbf{Z}_2\otimes\mathbf{Z}_2$ symmetry defined
by Eq. \eqref{eq:Z2trafo}. The corresponding order parameters are the two
real quantities 
\begin{align}
X_{1,2}=\langle \hat{\mathrm{X}}_{1,2}\rangle=\langle \hat{a}_{1,2}^{\phantom{\dag}}+\hat{a}_{1,2}^\dag\rangle=2\mathrm{Re}(\alpha_{1,2})\;.
\end{align}
A finite expectation value $X_i \neq 0$ creates the effective one-body potential $V_i$, which results in an atomic density wave. Thus equivalent order parameters can be defined by the density components
\begin{align}
\rho_{1,2}&=\int d\mathbf{r}\;\cos(\mathbf{k}_p\cdot\mathbf{r})\cos(\mathbf{k}_{1,2}\cdot\mathbf{r}) \langle \hat{\psi}^\dag(\mathbf{r}) \hat{\psi}(\mathbf{r})\rangle\notag\\&=\int d\mathbf{r}\;\cos(\mathbf{k}_p\cdot\mathbf{r})\cos(\mathbf{k}_{1,2}\cdot\mathbf{r}) \left(|\phi(\mathbf{r})|^2+\rho(\mathbf{r})\right)\;.
\end{align}
Additionally, we have the Bose-Einstein condensation transition
described by the $U(1)$ order parameter $\phi(\mathbf{r})$.

We first investigate the interplay
between the superradiant transition and the Bose-Einstein condensation
by solving the mean-field equations \eqref{eq:S_MF} as a function of
temperature $T$ and driving strength $U_p$. The corresponding phase diagram, which is qualitatively the same as the one for the single-cavity case considered in \cite{piazza_bose} is
shown in Fig.~\ref{fig:etaT_PD}\footnote{We choose $E_R=8\delta_c$ for our computations. This is much larger than the 
experimental values $E_R\sim\delta_c/100$ of the recoil energies, which would increase the numerical effort 
considerably without changing the qualitative physics.}. With growing values of the coupling strength $U_p$ the atomic gas becomes increasingly confined to the minima of the effective single-particle potential, which results in an enhancement of the kinetic energy and therefore a reduction of the critical temperature for Bose-Einstein condensation $T_c$. Beyond a critical coupling strength $U_{p_c}$ (or equivalently $\lambda_c$) the atoms spontaneously arrange into a spatially ordered configuration, resulting in a superradiant backscattering of light into the cavity. This Dicke-Hepp-Lieb transition as indicated by the blue line in Fig.~\ref{fig:etaT_PD} can be found both with and without a condensate
fraction. Additionally, a finite, but small, temperature can enhance the tendency
towards superradiance, as can be seen from the decrease of the
critical driving strength for increasing temperatures at $T\lesssim T_c^\text{ideal}$.

\begin{figure}[t]
\centering
\includegraphics [width=0.55\textwidth]{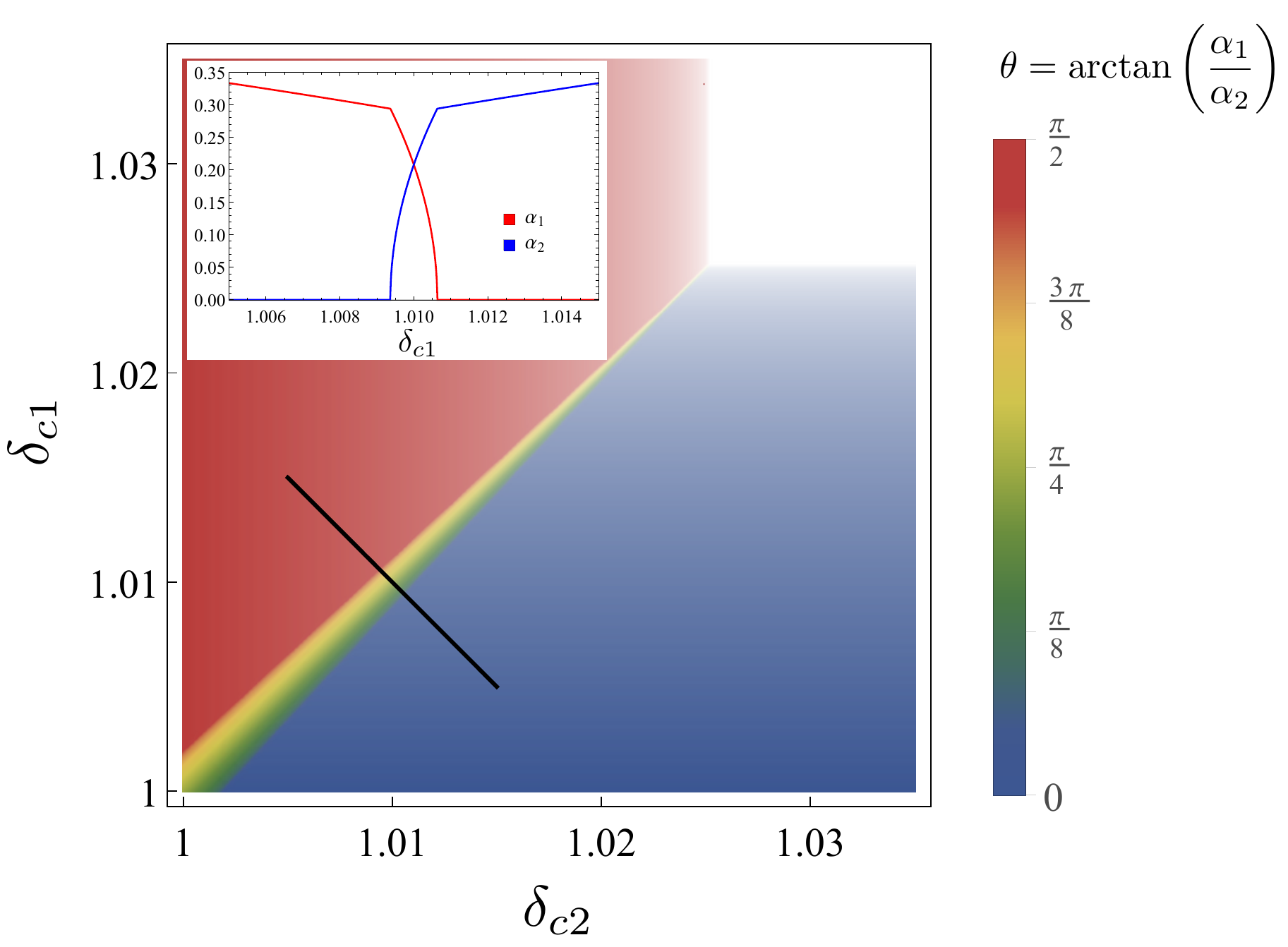}
\caption{Phase diagram in the $\delta_{c_1}-\delta_{c_2}$ plane. The color
  scale indicates the angle in $\theta$ in the $\alpha_1-\alpha_2$
  plane. Inset: cavity amplitudes along the black line indicated on
  the phase diagram. The parameters used are $T=0$, $\lambda=2.82$, $\eta=4$ and $\epsilon_\text{R}=8$.}
\label{fig:d1d2_PD}
\end{figure}

The phase diagram of Fig.~\ref{fig:etaT_PD} is computed for a
symmetric cavity configuration i.e. for equal detunings
$\Delta_1=\Delta_2$, implying that in the superradiant phase both cavities
are equally occupied with order parameters $\alpha_1=\alpha_2$.
For a comparison with the experimental results we also compute the 
zero temperature phase diagram in the $\delta_{c_1}-\delta_{c_2}$ plane, which is shown in
Fig.~\ref{fig:d1d2_PD}. Apart from the superradiant phases with only
one nonzero cavity field i.e. $\alpha_{1,2}\neq 0,\alpha_{2,1}=0$ we
observe a narrow region around the diagonal $\Delta_1=\Delta_2$
where both $\mathbf{Z}_2$ symmetries are broken i.e. $\alpha_{1}\neq
0,\alpha_{2}\neq 0$. Within this small region of the phase diagram the
two cavity order parameters are not equal, as quantified by the color
scale in Fig.~\ref{fig:d1d2_PD}, indicating the value of the angle
$\theta$ in the $\alpha_1-\alpha_2$ plane (see also Fig.~\ref{fig:eff_potential}), defined as
\[
\theta=\arctan\left(\frac{\alpha_1}{\alpha_2}\right)\;,
\]
which equals the $U(1)$-parameter of Eq.~\eqref{eq:U1trafo}.

This region exists due to the fact that the $U(1)$
symmetry of Eq.~\eqref{eq:U1trafo} is not perfectly realized. 
In particular, the size of the region is set by the value of the Goldstone mass.
Using Eq.~(\ref{eq:mG}) and the experimental parameters
$\Omega^2/|\Delta_A|=38 E_R$, $g^2/|\Delta_A|=5*10^{-4} E_R$ and $|\Delta|= 10^3 E_R$~\cite{leonard2016supersolid},
one obtains $|\Delta\epsilon|\sim m_G^2\alpha^2\sim 0.1\; E_R$, 
consistent with the experimental result $|\Delta\epsilon|< 10\; E_R$.

\section{Ginzburg-Landau potential for the cavity fields and role of cavity losses}
\label{sec:GLpotential}

\begin{figure}[t]
\centering
\includegraphics [width=0.7\textwidth]{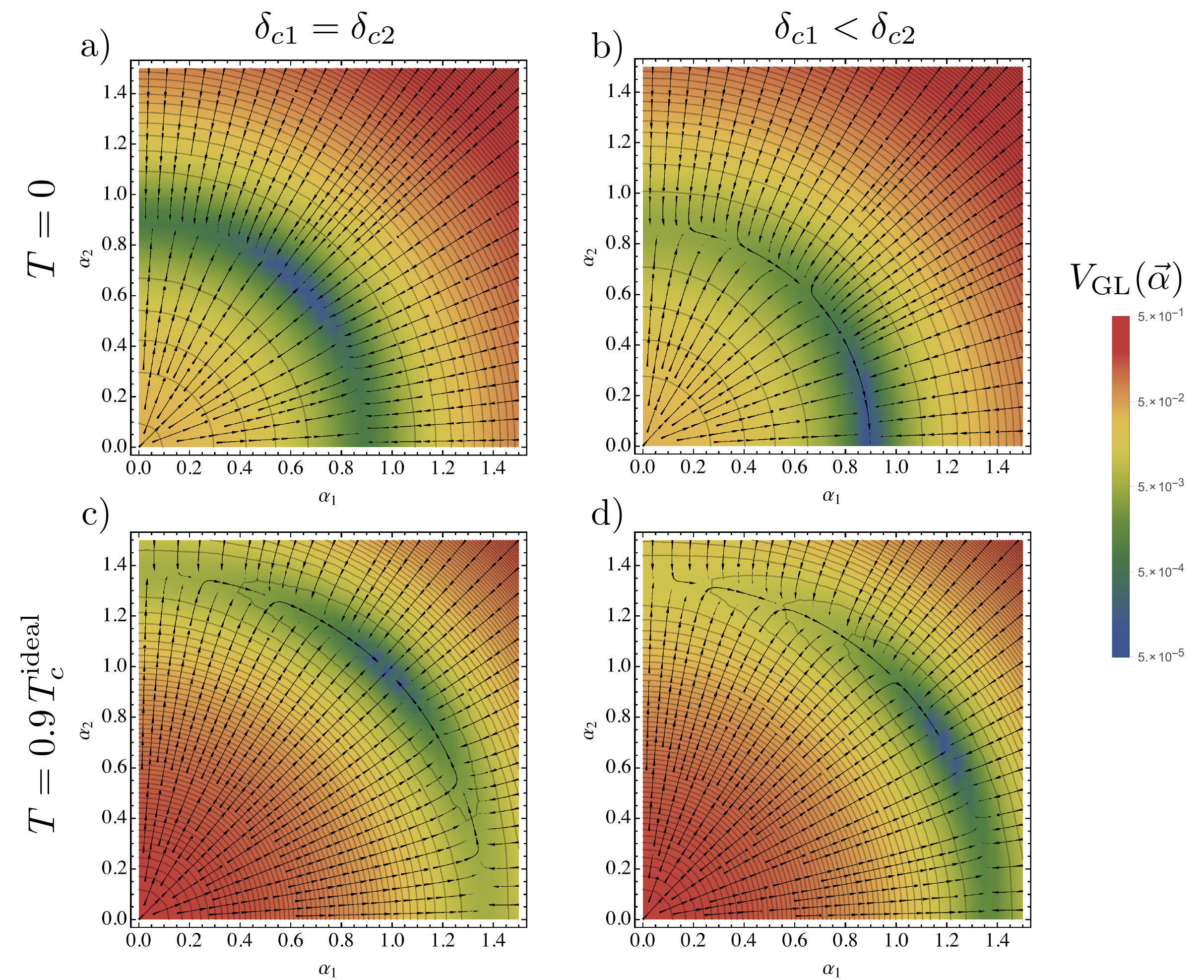}
\caption{Ginzburg-Landau potential $V_{\rm GL}$ as a function of the cavity
  amplitudes. Arrows indicate the gradient. a),b) are for $T=0$ while
  c),d) for $T=0.9 T_c^\text{ideal}$. a) and c) correspond to a symmetric
  configuration $\delta_{c_1}=\delta_{c_2}$, while b) and d) correspond
  to $\delta_{c_2}=1.01\delta_{c_1}$. The remaining parameters are
  $\eta=10$, $\epsilon_\text{R}=8$ as well as $\lambda=3$ in a), b) and $\lambda=2.7$ in c),d). Notice the small curvature along the azimuthal direction.}
\label{fig:eff_potential}
\end{figure}

In order to investigate the approximate $U(1)$ symmetry of our model in more detail,
we compute the full Ginzburg-Landau (GL) potential corresponding to the
mean-field Eqs.~\eqref{eq:mf}. The resulting effective potential in the
$\alpha_1-\alpha_2$ plane is shown in Fig.~\ref{fig:eff_potential}, 
both for an asymmetric and the perfectly symmetric choice of
detunings, at zero and finite temperature. 

The asymmetric case for $T=0$ is picked such that we are in the
single-cavity superradiant phase and the GL potential has indeed two minima at angles $\theta=0,\pi$ when cavity 1 is preferred,
or $\theta=\pm\pi/2$ when cavity 2 is preferred. The asymmetric case
for $T=0.9T_c$ is instead picked such that we still are in the
coexistence region where both cavities are occupied and where the GL
potential has four minima. One of those is shown at an angle slightly
below $\theta=\pi/4$, connected by reflection symmetry with respect to
the origin.

On the contrary, for $\Delta_1=\Delta_2$ the GL potential shows four degenerate
minima at $\theta=\pm\pi/4,\pm 3\pi/4$. Since a $U(1)$ symmetric
potential would show a degenerate minimum on a whole circle, we see
that the extent to which this symmetry is explicitly broken is
measured by the azimuthal curvature of the potential about anyone
of the four minima, which determines the square of the effective Goldstone mass. The latter,
together with the associated effective Higgs mass, which corresponds to the square root of the curvature in
the radial direction, is shown in Fig.~\ref{fig:masses} across the
superradiant phase transition at zero and at finite temperature.  
In the disordered phase there is only a single collective mode in the
radial direction with a mass vanishing at the critical point. Beyond
this point Goldstone and Higgs mass separate, the latter growing much
faster while the former remains at least one order of magnitude
smaller. By expanding the GL potential one can show that the Goldstone
mass close to the critical point is proportional to
$\sqrt{\alpha_1\alpha_2}$, in accordance with the arguments discussed in
section \ref{sec:symmetries}.
Moreover, the ratio between Goldstone mass and Higgs mass is inversely proportional to the drive
strength, so that for the strong drive employed in experiment and considered in
Fig.~\ref{fig:masses} we find a large separation between the Higgs and
the Goldstone mass. 

The qualitative behavior and the ratio of the Goldstone to Higgs mass shown in
Fig.~\ref{fig:masses} is consistent with the experimental results of
\cite{leonard_supersolid_goldstone}. By contrast, the presence of a
well-defined minimum in the GL potential in the range $0<\theta<\pi/2$ of
Fig.~\ref{fig:eff_potential} is not compatible with the experimental
finding \cite{leonard2016supersolid} that $\theta$ is homogeneously
distributed in this range. However, we can reconcile our prediction
with the experiment by adding the noise induced by cavity losses to
the picture. The probability of escaping the minimum and delocalizing across the circle in Fig.~\ref{fig:eff_potential} is given by $P_\text{deloc}\approx\exp(-\sqrt{N_{\rm ph}}\Delta V/\kappa)$, where $\Delta V\propto \alpha^4 \propto |\lambda-\lambda_c|^ 2$ is the depth
of the minimum while $\kappa$ is the cavity loss rate. Note that the noise is
suppressed by a factor $1/\sqrt{N_{\rm ph}}$ if we assume a coherent
cavity field.  With the experimental value $\kappa/2\pi\sim 200\,\text{kHz}$ and with $\Delta V\simeq |\Delta\epsilon|$ 
which is determined by the square of the Goldstone mass according to the first equality in Eq.~(\ref{eq:mG}),
typical values $N_{\rm ph}=|\alpha|^2\sim 100$ lead to an escape probability of $P_\text{deloc}\simeq\exp(-0.02)\simeq
0.98$. We stress that our estimate for $\Delta V$ is an upper bound
and thereby our escape probability provides a lower bound.
A critical test for this scenario of a restoration of the $U(1)$ symmetry by 
cavity loss induced noise, is that with an increasing number of
intracavity photons the escape probability is expected to decrease exponentially
like
\begin{align}
P_\text{deloc}\propto\exp(-N_{\rm ph}^{5/2})\;.
\end{align}

\section{Effective action for low-energy
  excitations}
\label{sec:low-energy-fluctuations}

In this final section we will discuss the nature of the low-energy
excitations of the cavity field in the superradiant phase. For this purpose, 
we expand the effective action derived in section \ref{sec:formalism} up to quadratic order
about the minima of the mean-field potential discussed in section~\ref{sec:GLpotential}. 
The resulting time-dependent deviations $a_i(\tau)$ can be expanded in terms
of discrete Fourier-coefficients  $a_{i,n}$ which determine the spectrum of 
light field fluctuations in the cavity. Thus, the effective Goldstone 
and Higgs mode appear explicitly, allowing 
to compute both their masses discussed above and - moreover - their damping 
or inverse lifetime which appears at finite temperature. \\

As described in section \ref{sec:formalism}, in the thermodynamic
limit the action \eqref{eq:S_eff} can be expanded up to quadratic order in the fluctuations. Since the coupling between atoms and the imaginary part of the cavity fields results solely in a dispersive shift, we can integrate out the imaginary part, generating only even powers in $\omega_n$. At zero temperature the fluctuation part in dimensionless units is then given by
\begin{align}
\label{eq:Sfluct_T=0}
S_\text{eff}&^{(FL)}[a^R_{1,2}]\bigg|_{T=0}=\sum_{n\neq0}\sum_{i=1,2}\Bigg\{\left(\omega_n^2+1\right)a^R_{i,n}a^R_{i,-n}\notag\\
&\left.+4\sum_{j=1,2}\sum_l\frac{1}{i\omega_n-\epsilon_l(0)}\langle\Psi_0(0)|\frac{\partial
      V_\text{sp}}{\partial \alpha_i^*}|\Psi_l(0)\rangle
    \langle\Psi_l(0)|\frac{\partial V_\text{sp}}{\partial
      \alpha_{j}^*}|\Psi_0(0)\rangle a^R_{i,n}a^R_{j,-n}\right\}\;,
\end{align}
where $a_i^R$ is a real part of the cavity field and $|\Psi_l(k)\rangle$ is the atomic wave function with quasi-momentum $k$ and band index $l$. This expression describes the scattering of atoms from the condensate to the $\Gamma$-point of an excited band in second order perturbation theory. 
Since these processes are far off-resonant with
respect to the low energy excitations in the photon fields, they do not give rise to damping. The associated 
spectral functions are thus perfectly sharp. The picture gradually changes with
increasing temperature, when more and more atoms occupy states near
the edge of the Brillouin zone, where low energetic photons can be
scattered resonantly. This effect can be accounted for by generalizing the effective action through the inclusion of thermally occupied states
\begin{align}
\label{eq:Sfluct}
&S_\text{eff}^{(FL)}[a^R_{1,2}]=n_0 S_\text{eff}^{(FL)}[a^R_{1,2}]\bigg|_{T=0}\nonumber\\
&\!+\!\!\sum_{j=1,2}\sum_{m,l} \!V\!\!\int\!\frac{d^3
      k}{4\pi^3}\langle\Psi_m(k)|\frac{\partial
      V_\text{sp}}{\partial \alpha_i^*}|\Psi_l(k)\rangle
    \langle\Psi_l(k)|\frac{\partial V_\text{sp}}{\partial
      \alpha_{j}^*}|\Psi_m(k)\rangle\frac{n_b(\epsilon_m(k))\!-\!n_b(\epsilon_l(k))\!}{i\omega_n+\epsilon_m(k)-\epsilon_l(k)}a^R_{i,n}a^R_{j,-n}\nonumber\\
& \equiv\sum_{n\neq0}\sum_{i,j}\mathcal{G}(\omega_n)_{i,j}a^R_{i,n}a^R_{j,-n}\;,
\end{align}
where $n_0$ is the condensate fraction. An important point is that, both at $T=0$ and at finite temperature, 
the action involves only the real parts of the cavity fields and is thus an even function of $\omega$. Therefore,
it contains no linear terms of the form $i\hbar a^R(t)\partial_t a^R(t)$ which would give rise to dynamics 
involving a reversible first order time derivative, where no proper Higgs mode exists~\cite{pekker2015}. 
Since the matrix elements respect the symmetry of the mean field
action, the fluctuations can be diagonalized in terms of Goldstone
and Higgs modes $a_G=-\sin{\theta}a_1^R+\cos{\theta} a_2^R$ and
$a_H=\cos{\theta}a_1^R+\sin{\theta} a_2^R$. 
Upon expanding to second order in the frequency, we thus obtain the action
\begin{align}
S_\text{eff}^{(FL)}[a_G,a_H]\approx\left(Z_G\omega_n^2+ m_G^2\right)a_{G,n}a_{G,-n}+\left(Z_H\omega_n^2+ m_H^2\right)a_{H,n}a_{H,-n}
\end{align}
with numerical coefficients that fulfill $m_G \ll m_H$, as well as
$Z_{G,H}-1=\mathcal{O}(m_{G,H}/\epsilon_R)$ at small
temperatures. From this action the
existence of a (approximately) gapless Goldstone mode together with a
strongly gapped Higgs mode is apparent.

\begin{figure}[htp]
\centering
\includegraphics [width=0.6\textwidth]{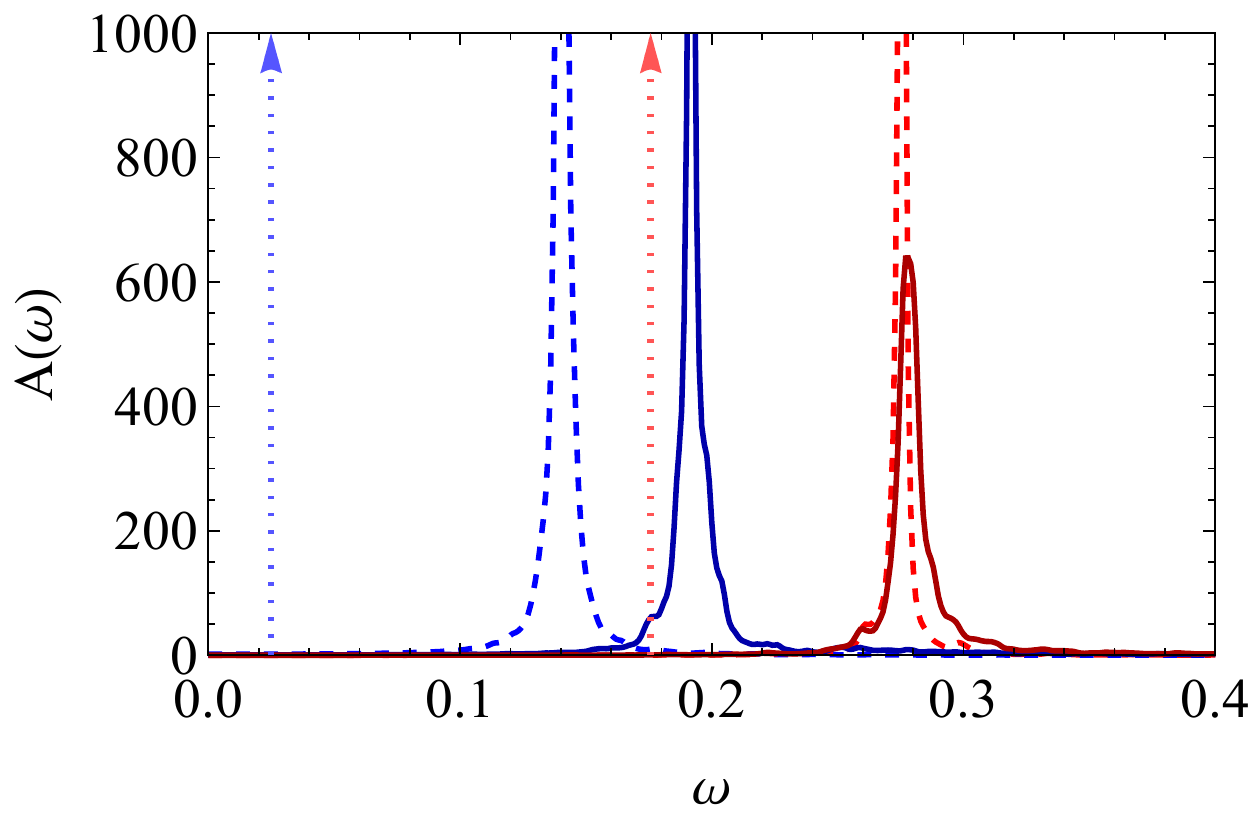}
\caption{Low-energy excitation spectrum in the $\eta-\omega$ plane
  showing the Goldstone and Higgs peak with finite width close to the critical point. The temperatures and coupling strengths $\lambda$ in order of increasing color saturation are $\left\{T,\lambda\right\}=\{0,2.2\}$ (dotted arrows), $\{0.5,2.3\}$ (dashed line) and $\{0.9,2.2\}$ for the solid line.}
\label{fig:spectrum}
\end{figure}

As anticipated, at finite temperatures the Goldstone and Higgs modes experience losses via resonant
Landau damping processes where a photon scatters against an atom while conserving energy
and momentum.
The resulting lifetimes as well as the masses of both
excitations can be obtained from the spectral function $A(\omega)=2
\Im{\mathcal{G}(-i\omega+0^+)}$ which can be measured via
pump-probe experiments. The resulting spectra are shown in Fig.~\ref{fig:spectrum}
for different temperatures \footnote{Note that up to leading order in
  the frequency expansion the lifetime of the modes is infinite and we
need to use the full action (\ref{eq:Sfluct}) in order to introduce
damping.}. 
For finite temperatures, there is additional structure in
the tails of the Goldstone and Higgs peaks, which arises from van-Hove
 singularities at the edges of the Brillouin zones.

\begin{figure}[t]
\centering
\includegraphics [width=0.8\textwidth]{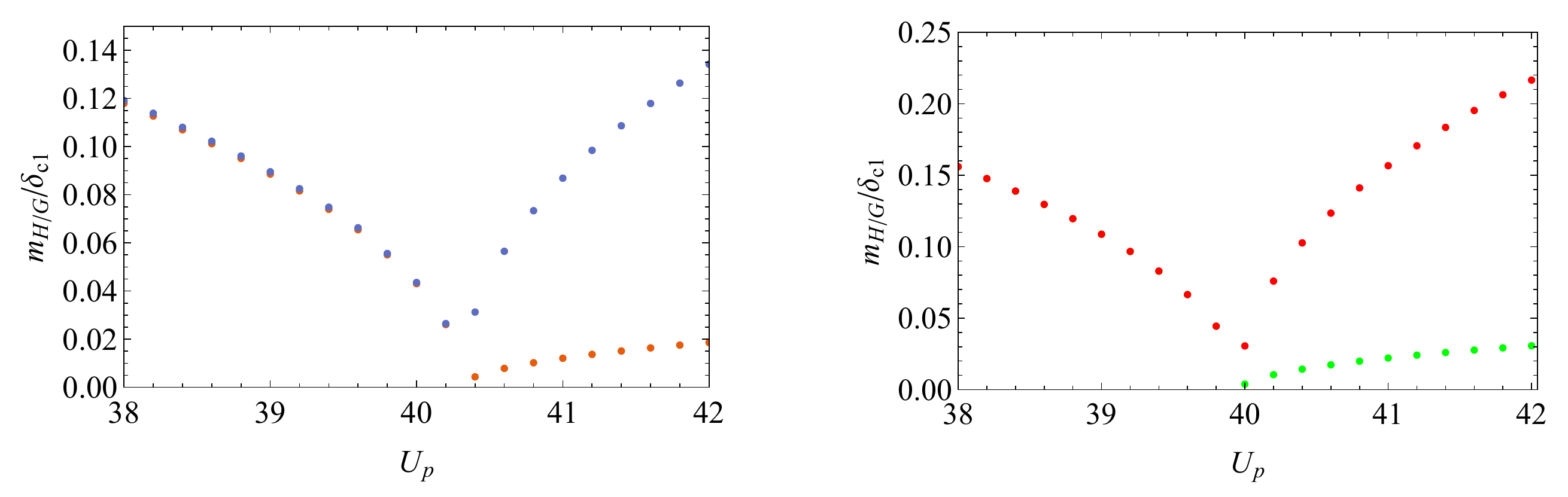}
\caption{Goldstone and Higgs mass across the superradiant
  transition at $T=0$(left) and $T=0.5 T_c^\text{ideal}$ (right). Parameters are the same as in Fig.~\ref{fig:etaT_PD} apart from $\lambda=2.278$.}
\label{fig:masses}
\end{figure}

As shown in Fig.~\ref{fig:spectrum}, the Goldstone mass increases
drastically with temperature, an effect that cannot be observed in
Fig.~\ref{fig:masses} for the mass obtained from the curvature of the mean field
action at the global minimum. This is because in our expansion in fluctuations about the
potential minima we do not allow atoms to redistribute. We 
are therefore effectively computing the behavior of photonic excitations at "high" frequencies
with respect to the timescale of atomic redistribution. 
The inclusion of atomic redistribution beyond mean field would require
a non-equilibrium approach like the one employed in
\cite{piazza_QKE}. This would allow to interpolate smoothly between
the high frequency mass, as determined in $A(\omega)$, and low
frequency mass, obtained from the mean-field potential.
However, since the atomic redistribution time is extensive in the number of atoms \cite{piazza_QKE}, we 
expect the Goldstone mass experimentally observable in large systems to
correspond to the high-frequency mass measured by the spectral function.\\

The spectral function exhibits two distinct peaks which possess a nontrivial frequency but no momentum dependence. 
This is a consequence of the fact that the present double-cavity system is still an effectively zero-dimensional one. 
As a result, it does not give rise to a genuine spectrum of Goldstone modes usually associated with supersolids, 
where gauge and translation symmetry are broken simultaneously in a system with short range interactions. 
For such a genuine supersolid, the total free energy can be written as an integral over a spatially varying free
energy density $f(T,n,\mathbf{v}_s,\underline{u})$ which involves thermodynamic variables which vary continuously in space. 
In particular, the simultaneous presence of broken translation and gauge invariance leads to 
two additional contributions in the differential of the free energy density
\begin{align}
df\big|_{T,n}=\mathbf{j}_s\cdot d\mathbf{v}_s+{\rm Tr}\, \underline{\sigma}\cdot d \underline{u}\;.
\end{align}
As discussed by Liu within a hydrodynamic approach~\cite{liu1978}, the term proportional to the superfluid 
current density $\mathbf{j}_s$ and its conjugate variable, the  
superfluid velocity $\mathbf{v}_s=\frac{\hbar}{m}\nabla\phi$, results in a persistent mass flow for
generic supersolids or dissipationless entropy flow in the absence of defects. 
Similarly, for the generic case of short range interactions where the stress tensor $\underline{\sigma}$ 
is linearly proportional to the strain tensor $\underline{u}$,
the second contribution gives rise to phonons whose frequency $\omega(\mathbf{q})\sim |\mathbf{q}|$ vanishes linearly with 
the wave vector. Due to  $\omega(\mathbf{q})=\omega(\mathbf{q}+\mathbf{G})$ for regular crystals, this entails
a vanishing Goldstone mass $\omega(\mathbf{q}=\mathbf{G})=0$ at reciprocal lattice vectors as a signature
of the spontaneous breaking of translation invariance~\cite{wagner1966}. In the present system such a Goldstone mode 
also exists for the motion of atoms in the limit where $m_G$ can be neglected. 
It is associated with the shift along the $x$-direction discussed in Eq.~\eqref{eq:U1trafo} and leads to $\omega(\mathbf{G})=0$ for all reciprocal lattice vectors $\mathbf{G}=n\mathbf{k}_1+m\mathbf{k}_2$ with $n,m \in \mathbb{Z}$. In particular, the transverse acoustic phonon at $n=-m=\pm1$ corresponds to the translation described in Eq.~\eqref{eq:U1trafo}, which is related to the indirect exchange of a photon between the two cavities.
In contrast to the standard situation, however, where the phonon frequencies approach zero continuously as $\mathbf{q}$ approaches $0$, the long ranged nature of the interactions give rise to a finite energy gap at any $\mathbf{q}\neq\mathbf{G}$. The Goldstone mode thus exists only at isolated points in momentum space, with all other momenta being gapped.

\section{Conclusions}

In summary, we have studied the nature of broken symmetries, the effective Ginzburg-Landau potential and the 
spectrum of the light field in the double cavity setup realized recently at ETH~\cite{leonard2016supersolid,leonard_supersolid_goldstone}. 
It has been shown that the emergent $U(1)$ 
invariance for symmetrically coupled cavities is slightly broken by higher order photon scattering processes. 
We have determined an upper bound for the resulting mass of the effective Goldstone mode which is consistent with the experimental 
results~\cite{leonard_supersolid_goldstone}. In addition, it has been shown
that the ratio $m_G/m_H$ between the Goldstone and Higgs mass vanishes in the limit of large driving amplitudes. 
As an experimentally testable prediction, we have determined the cavity noise induced escape probability from the global minimum 
of the effective potential as a function of the intracavity photon occupation which might be used for an indirect measurement of the Goldstone mass.
Finally, the issue of dissipationless transport of particles in the double cavity supersolid has been discussed carefully and
has been compared to the case of genuine supersolids, where this is associated with an additional true Goldstone mode.

Acknowledgments: W. Z. would like to thank the Quantum Optics group at ETH for the great hospitality and for
numerous discussions during a sabbatical, where this work has been started. We are grateful to Andrea Morales and Julian L{\'e}onard for careful reading and comments on the manuscript.  

\section*{References}

\bibliography{mybib}

\end{document}